\documentclass[twocolumn,showpacs,preprintnumbers,amsmath,amssymb]{revtex4}

\newcommand{\commentoutA}[1]{}

\usepackage{graphicx}% Include figure files
\usepackage{dcolumn}% Align table columns on decimal point
\usepackage{bm}% bold math
\begin{document}

\preprint{LA-UR 07-7769}

%\title{Lagrangian formulation of time-reversible Born-Oppenheimer molecular dynamics for 
%higher-order symplectic integration}
\title{Extended Born-Oppenheimer molecular dynamics}

\author{Anders M. N. Niklasson}
\affiliation{Theoretical Division, Los Alamos National Laboratory, Los Alamos, New Mexico 87545}
\affiliation{Applied Materials Physics, Department of Materials Science and
Engineering, Royal Institute of Technology, SE-100 44 Stockholm, Sweden}

\date{\today}

\begin{abstract}
A Lagrangian generalization of time-reversible Born-Oppenheimer molecular dynamics
[Niklasson {\em et al.}, Phys. Rev. Lett. {\bf 97}, 123001 (2006)] is proposed.
The Lagrangian includes extended electronic degrees of freedom as auxiliary 
dynamical variables in addition to the nuclear coordinates and momenta. 
While the nuclear degrees of freedom propagate on the Born-Oppenheimer potential
energy surface, the extended auxiliary electronic degrees of freedom evolve as a harmonic
oscillator centered around the adiabatic propagation of the self-consistent ground state.
The formulation enables the application of higher-order symplectic or geometric
integration schemes that are stable and energy conserving even under incomplete
self-consistency convergence.  It is demonstrated how the extended Born-Oppenheimer 
molecular dynamics improves the accuracy by over an order of magnitude compared 
to previous formulations at the same level of computational cost. 
\end{abstract}


%\keywords{electronic structure theory, molecular dynamics, Born-Oppenheimer
%molecular dynamics, Car-Parrinello molecular dynamics, symplectic, time-reversal symmetry, lossless, density
%functional theory, Hartree-Fock, density matrix theory, detailed balance, linear scaling
%electronic structure theory, hysteresis, energy conserving, phase-space conserving, ab initio molecular dynamics, 
%extrapolation, energy conservation, time-reversibility, energy drift, microcanonical, stability, incomplete convergence,
%lossless filter, self-consistent field, perfect reconstruction, extended Lagrangian, Euler-Lagrange, equations of motion}
\maketitle

%\section{Introduction}

Born-Oppenheimer molecular dynamics \cite{CLeforestier,THelgaker90,RBarnett91,RWentzcovitch91,MCPayne92,DMarx00}
based on self-consistent field (SCF) methods, such as Hartree-Fock or density functional theory \cite{Roothaan,hohen,KohnSham65},
is currently a gold standard in molecular dynamics simulations.
It is derived from the well defined adiabatic approximation for the separation of the nuclear and electronic
degrees of freedom, where the forces acting on the atoms are calculated at the self-consistent electronic ground state \cite{PPulay69}.
However, the ability to achieve physically accurate and stable microcanonical simulations, while keeping
the computational cost low, has been limited by the requirement of a high degree of SCF convergence
in the nuclear force calculations \cite{PPulay04,JMHerbert05}. Only very recently,
by restoring the time-reversal symmetry in the underlying adiabatic propagation of
the electronic degrees of freedom \cite{JKolafa03,ANiklasson06,TDKuhne07}, 
has it been possible to achieve efficient energy conserving simulations 
also under incomplete SCF convergence. Unfortunately, these techniques
are not able to take advantage of powerful symplectic or
geometric integration methods developed for celestial and classical molecular dynamics
\cite{RRuth83,Feng86,HYoshida90,MTuckerman92,DFrenkel02,BLeimkuhler04}.
Symplectic integration, which in general requires a Hamiltonian or Lagrangian formulation
of the dynamics, enables highly efficient simulations while keeping a rigorous
control over physical properties.

The purpose of this Letter is to take advantage of geometric integration methods in Born-Oppenheimer 
molecular dynamics by introducing a Lagrangian generalization of the recently proposed time-reversible 
Born-Oppenheimer molecular dynamics \cite{ANiklasson06}. This gives time-reversible 
Born-Oppenheimer molecular dynamics a theoretically more solid and physically transparent framework
and, most significantly, thanks to the Lagrangian formulation, higher-order symplectic
integration algorithms can be applied, which provide superior performance in molecular dynamics simulations.
It will be demonstrated how the accuracy of the extended Born-Oppenheimer molecular dynamics is increased 
by over an order of magnitude at the same level of computational cost compared to previous formulations.

The conventional Born-Oppenheimer (BO) Lagrangian for {\em ab initio} molecular
dynamics is given by
\begin{equation}\label{LG}
{\cal L}^{\rm BO}({\bf R},{\bf \dot R}) = \frac{1}{2} \sum_k M_k {\dot R}_k^2 - U_{\rm SCF}[{\bf R};D],
\end{equation}
where the potential $U_{\rm SCF}$ is the total electronic energy in, for example, Hartree-Fock
or density functional theory, including the nuclear-nuclear repulsion terms. The potential energy 
$U_{\rm SCF}[{\bf R};D]$ is defined at the electronic ground state given by the 
SCF optimized solution $D$ of the electronic degrees of freedom.
$D$ is assumed to be the symmetric single-particle density matrix in an orthogonal
basis-set representation, though generalizations to other representations,
such as the density, the wavefunctions, or the Kohn-Sham Hamiltonian, are straightforward.
Notice that $D$ is not an independent dynamical variable, since 
it is determined by the external potential at atomic configuration ${\bf R} = \{R_k\}$. It
is included to show that the Lagrangian ${\cal L}^{\rm BO}$ is calculated 
at the self-consistent Born-Oppenheimer ground state.
The nuclear degrees of freedom are given by the atomic coordinates $R_k$ and velocities 
${\dot R}_k$, with the corresponding masses $M_k$.  The dots denote time derivatives.

The key result of this Letter is the extension of the Born-Oppenheimer Lagrangian 
${\cal L}^{\rm BO}$ in Eq.\ (\ref{LG}) by the addition of auxiliary electronic degrees 
of freedom $P$ and ${\dot P}$ that evolve in a harmonic potential centered around the 
self-consistent solution $D$.  The extended auxiliary dynamical variables, $P$ and ${\dot P}$, 
are assumed to be of the same form as $D$, i.e. a density matrix and its time derivative.
The extended Born-Oppenheimer (XBO) Lagrangian is given by
\begin{equation}\label{LN}
{\cal L}^{\rm XBO}({\bf R},{\bf \dot R},P,{\dot P}) = {\cal L}^{\rm BO} +
\frac{\mu}{2} Tr[{\dot P}^2] - \frac{\mu \omega^2}{2} Tr[(D-P)^2].
\end{equation}
Here $\mu$ and $\omega$ are fictitious mass and frequency parameters
for the auxiliary electronic degrees of freedom.  There are no additional terms imposing constraints
on the electronic degrees of freedom, i.e.\ wavefunction orthogonality 
or density matrix idempotency \cite{RCar85,HBSchlegel01,JMHerbert04}.  These constraints are not necessary, 
since the potential energy $U_{\rm SCF}[{\bf R};D]$ and the nuclear forces are calculated at the 
normalized and idempotent ground state $D$. 

The time evolution of the dynamical system described by the extended Lagrangian
${\cal L}^{XBO}$ is determined by Euler-Lagrange equations of motion:
\begin{equation}
%M_k {\ddot R}_k = \frac{\partial {\cal L}^{XBO}}{\partial R_k} = 
%\frac{\partial {\cal L}^{BO}}{\partial R_k} = -\frac{\partial U_{\rm SCF}\left[{\bf R};D\right]}{\partial R_k},
M_k {\ddot R}_k = 
-\frac{\partial U_{\rm SCF}\left[{\bf R};D\right]}{\partial R_k} - \mu \omega^2 Tr[(D-P) \partial D/\partial R_k],
\end{equation}
\begin{equation}
%\mu {\ddot P} = \frac{\partial {\cal L}^{XBO}}{\partial P} = \mu {\omega^2} (D - P).
\mu {\ddot P} = \mu {\omega^2} (D - P).
\end{equation}
In the limit $\mu \rightarrow 0$, i.e. when ${\cal L}^{XBO} \rightarrow {\cal L}^{BO}$, the dynamics
is determined by the equations of motion,
\begin{equation}\label{EL_1}
M_k {\ddot R}_k = -\frac{\partial U_{\rm SCF}\left[{\bf R};D\right]}{\partial R_k},
\end{equation}
\begin{equation}\label{EL_2}
{\ddot P} = {\omega^2} (D - P).
\end{equation}
These two equations of motion reveal two properties that are of fundamental importance 
in the extended Born-Oppenheimer molecular dynamics:
$a$) The nuclear forces are calculated at the self-consistent ground state $D$
as with the Born-Oppenheimer Lagrangian. The molecular trajectories therefore evolve 
on the Born-Oppenheimer potential energy surface with the total Born-Oppenheimer energy,
$E^{\rm BO} =  \frac{1}{2} \sum_k M_k {\dot R}_k^2 + U_{\rm SCF}[{\bf R};D]$,
as a constant of motion.
$b$) The equations of motion do not include the fictitious electron mass 
parameter $\mu$, which otherwise could cause problems \cite{PTangney06}.

Since the extended electronic degrees of freedom $P(t)$ evolve in a harmonic
potential centered around their own self-consistent solutions $D(t)$ in Eq.\ (\ref{LN}), the auxiliary density matrix $P(t)$ and
its self-consistent solution $D(t)$ will stay close together. We can therefore use $P(t)$ as an efficient initial guess to $D(t)$
in the iterative SCF optimization,
\begin{equation}\label{SCFOP}
D(t) = {\rm SCF} [{\bf R}(t),P(t)].
\end{equation}
This strongly reduces the computational cost to reach the
self-consistent ground state $D(t)$ at which the nuclear forces are calculated in
Eq.\ (\ref{EL_1}).  

In conventional Born-Oppenheimer molecular dynamics the initial guess for the
iterative SCF optimization is not given by an auxiliary dynamical variable, as in Eq.\ (\ref{SCFOP}), 
but by an extrapolation from previous time steps \cite{MCPayne92,TAArias92,JMMillan99,PPulay04,CRaynaud04,JMHerbert05}. 
Unfortunately, because the SCF procedure is irreversible and in practice never complete, 
this extrapolation breaks the time-reversal symmetry in the underlying propagation of the electronic
degrees of freedom, which causes serious stability problems with a systematic drift in
the total energy \cite{PPulay04,JMHerbert05}.  Only by increasing the SCF convergence, 
at great computational cost, can the energy drift be reduced, though it never fully disappears.
The fundamental problem with the broken time-reversal symmetry in the electron propagation was recently solved by 
the introduction of time-reversible Born-Oppenheimer molecular dynamics
based on a lossless dual filter integration scheme \cite{ANiklasson06}.
With the extended Lagrangian formulation the time-reversal problem is avoided in a similar way: 
the auxiliary electronic degrees of freedom $P(t)$, and thus the initial SCF guesses in Eq.\ (\ref{SCFOP}),
occur, not through extrapolation, but as dynamical variables that can be integrated by time-reversible 
algorithms \cite{LVerlet67,WCSwope82,DFrenkel02}.
The nuclear forces are then calculated with an underlying electron propagation that is time reversible.
In this way a systematic energy drift is avoided even under incomplete SCF convergence.

It is easy to see that the extended Lagrangian formulation is a generalization of time-reversible
Born-Oppenheimer molecular dynamics.  If we apply the time-reversible Verlet scheme \cite{LVerlet67}
to the integration of the electronic degrees of freedom in Eq.\ (\ref{EL_2}) we get
\begin{equation}\label{TRL}
P(t + \delta t) = 2P(t) - P(t-\delta t) + \delta t^2 \omega^2\big(D(t)-P(t)\big).
\end{equation}
If we choose the dimensionless factor $\kappa = \delta t^2 \omega^2 = 2$ this propagation 
is identical to the linear integration scheme in time-reversible 
Born-Oppenheimer molecular dynamics \cite{ANiklasson06}.
Thus, the extended Born-Oppenheimer Lagrangian in Eq.\ (\ref{LN})
forms a natural framework for time-reversible Born-Oppenheimer molecular dynamics, which 
thereby is given a more rigorous and physically transparent formulation. Instead
of propagation through a time-reversible dual filter process, the auxiliary electronic
degrees of freedom $P(t)$ occur as dynamical variables that evolve through
a time-reversible integration scheme.

Possibly the most important advantage of the extended Lagrangian formulation of time-reversible
Born-Oppenheimer molecular dynamics is that it enables the application of higher-order
symplectic or geometric integration methods 
\cite{RRuth83,Feng86,HYoshida90,MTuckerman92,DFrenkel02,BLeimkuhler04}.
Whereas a conventional integration algorithm can be seen as a numerical approximation for 
the integration of an underlying exact Hamiltonian dynamics, a symplectic integration 
can be seen as an exact integration for an underlying approximate Hamiltonian. 
The conservation of various physical properties of the approximate Hamiltonian can then 
be rigorously controlled \cite{HYoshida90,McLachlan92,BLeimkuhler04}.  
For the nuclear coordinates in Eq.\ (\ref{EL_1})
a quite general symplectic integration \cite{HYoshida90,McLachlan92}
over a time length $\delta t$ is divided in $m$ steps $(i = 1,2,\ldots, m)$
\begin{equation}\begin{array}{l}\label{SR}
{\dot R}_k(t_i) = {\dot R}_k(t_{i-1}) + b_i \delta t {\ddot R}_k(t_{i-1}),\\
R_k(t_i) = R_k(t_{i-1}) + a_i \delta t {\dot R}_k(t_i).
\end{array}
\end{equation}
Here $[R_k(t_0),{\dot R}_k(t_0)] = [R_k(t), {\dot R}_k(t)]$ and 
$[R_k(t + \delta t),{\dot R}_k(t + \delta t)] = [R_k(t_m),{\dot R}_k(t_m)]$. 
For the electronic degrees of freedom in Eq.\ (\ref{EL_2}),
for $i = 1,2,\ldots, m$, and using the variable substitution $\delta t {\dot P}(t) \rightarrow {\dot P}(t)$, 
the symplectic integration is
\begin{equation}\label{SI}\begin{array}{l}
{\dot P}(t_i) = {\dot P}(t_{i-1}) +  b_i \kappa \left(D(t_{i-1}) - P(t_{i-1})\right),\\
P(t_i) = P(t_{i-1}) +  a_i {\dot P}(t_i),
\end{array}
\end{equation}
where $D(t_{i}) = {\rm SCF}[{\bf R}(t_i),P(t_i)]$.
Here $\kappa = \delta t^2 \omega^2$, $[P(t_0),{\dot P}(t_0)] = [P(t), {\dot P}(t)]$ and
$[P(t + \delta t),{\dot P}(t + \delta t)] = [P(t_m),{\dot P}(t_m)]$.
Examples of coefficients $a_i$ and $b_i$ for various number of steps $m$ can 
be found in Ref.\ \cite{McLachlan92}. 

Before applying a symplectic integration algorithm the value of the dimensionless constant 
$\kappa = \delta t^2 \omega^2$ must be chosen.  Since the SCF convergence is always incomplete
the self-consistent solutions $D(t_i)$ will be calculated only approximately.
The idea is to choose $\kappa$ such that the integration in Eq.\ (\ref{SI}) is always stable under approximate SCF convergence.
The optimal choice is the largest $\kappa$-value that is consistent with stability, since this choice gives
the largest value of $w^2$ for a given time step $\delta t$. A larger $w^2$ corresponds to a higher 
curvature of the harmonic potential in Eq.\ (\ref{LN}), which keeps the 
auxiliary dynamical variables closer to the self-consistent ground state.  This reduces the error 
and/or the cost for the SCF optimization.  Based on a linearization of the SCF optimization procedure 
in Eq.\ (\ref{SCFOP}) around its exact ground state $D^*$, we can express an approximately 
SCF optimized density matrix as
\begin{equation}\label{SCFLin}
D \approx D^* + \Gamma \left( P-D^* \right).
\end{equation}
Here $\Gamma$ corresponds to the SCF response kernel, which is given as a ``super matrix'' acting
on the matrix $\left( P-D^* \right)$. Assuming at least some amount of convergence
in the SCF procedure the eigenvalue of $\Gamma$ with the largest magnitude, $\gamma$, will be
somewhere in the interval $\gamma \in[-1,1]$. Following the analysis by Arias {\em et al.} \cite{TAArias92},
we insert the linearized SCF expression in Eq.\ (\ref{SCFLin}), with $\Gamma$ replaced by $\gamma$,
in the symplectic integration, Eq.\ (\ref{SI}), and look at the homogeneous part of the equation,
\begin{equation}\label{Stab}
\left[\begin{array}{l} {\dot P}(t)\\ P(t)\\ \end{array} \right] = {\bf T}_m {\bf T}_{m-1} \ldots {\bf T}_1
\left[\begin{array}{l} {\dot P}(t-\delta t)\\ P(t-\delta t)\\ \end{array} \right].
\end{equation}
Here ${\bf T}_i$ ($i = 1,2,\ldots, m$) are the matrices
\begin{equation}\label{TMat}
{\bf T}_i = \left[\begin{array}{ccc} 1 & b_i \kappa (\gamma-1)\\ a_i & a_ib_i \kappa (\gamma-1)+1 \end{array} \right].
\end{equation}

Equation (\ref{Stab}) corresponds to a mapping of the phase space from one time step to the next for a linearized
test system with the constant solution $D^*(t) \equiv 0$.
The optimal choice of $\kappa$ is the largest value for which the mapping ${\bf T}_n {\bf T}_{n-1} \ldots {\bf T}_1$
has all its eigenvalues on the unit circle for all degrees of incomplete SCF convergence, i.e.\ for $\gamma \in [-1,1]$.
This case avoids exponentially increasing solutions leading to numerical instabilities or unphysical dissipation. 
The mapping in Eq.\ (\ref{Stab}) always preserves the phase space, i.e. the ``area'' spanned by ${\dot P}(t)$ 
and $P(t)$, since the determinant of the mapping 
$\det({\bf T}_m {\bf T}_{m-1} \ldots {\bf T}_1) = \Pi_{i=1}^{m} \det({\bf T}_i) = 1$ 
for all values of $\kappa$ and $\gamma$.

\begin{figure}[t]
\resizebox*{3.0in}{!}{\includegraphics[angle=00]{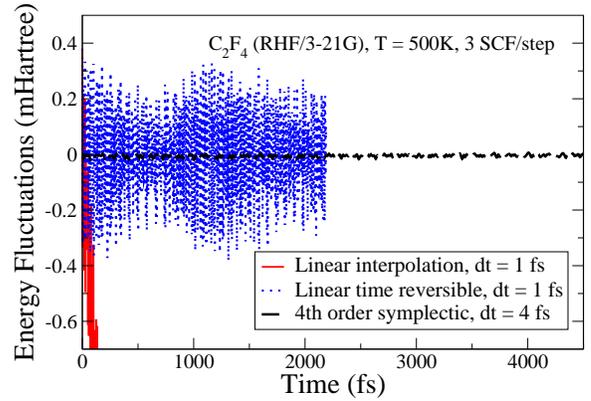}}
\caption{\label{Fig1}
The fluctuations in total Born-Oppenheimer energy ($E^{\rm BO}(t)-E_0$) using three different
{\em ab initio} molecular dynamics approaches described in the text. The integration 
time length $dt$ is adjusted to allow for a direct comparison between the methods at the same level of computational cost.}
\end{figure}

For the optimal 4th order integration scheme by McLachlan and Atela \cite{McLachlan92} 
(where 
$a_1 \approx 0.515352837$,
$a_2 \approx -0.0857820194$,
$a_3 \approx 0.441583024$,
$a_4 \approx 0.128846158$,
$b_1 \approx 0.134496199$,
$b_2 \approx -0.224819803$,
$b_3 \approx 0.756320001$, and
$b_4 \approx  0.334003603$) 
the largest possible $\kappa$-value consistent
with stability under incomplete SCF convergence is $\kappa = 4.617$.  
For the conventional Leap-Frog or Velocity Verlet scheme \cite{McLachlan92}, 
as well as the time-reversible Verlet integration in Eq.\ (\ref{TRL}),
the largest possible value is $\kappa = 2$.

The great advantage with stability under incomplete SCF convergence is that any amount
of convergence suffices for stability, which typically means that only one SCF cycle
per force calculation is necessary. This is in contrast to possibly all previous higher-order 
extrapolation schemes (beyond linear order), for example, the Fock-Matrix dynamics 
schemes by Pulay and Fogarasi \cite{PPulay04} and the higher-order 
time-reversible schemes by Niklasson {\em et al.} \cite{ANiklasson06}, which are
unstable under incomplete SCF convergence, i.e.\ for the full interval of $\gamma \in [-1,1]$.  
Simulations based on those methods will thus diverge if not a certain finite 
degree of SCF convergence can be guaranteed.
To take full advantage of the higher-order symplectic integration schemes the accuracy in
the electronic force calculations in Eqs.\ (\ref{EL_2}) and (\ref{SI}) should preferably match the accuracy
in the integration of the nuclear degrees of freedom. This generally motivates an improved SCF convergence,
though it is not a requirement for stability.

Figure \ref{Fig1} shows the fluctuations in the total Born-Oppenheimer energy  $E^{\rm BO}$
%($E^{\rm BO} =  \frac{1}{2} \sum_k M_k {\dot R}_k^2 + U_{\rm SCF}[{\bf R};D]$)
for a C$_2$F$_4$ molecule using
Hartree-Fock theory with a standard Gaussian basis set for three schemes: $i)$ conventional linear
interpolation of the electronic degrees of freedom from two previous time steps, $ii)$ time-reversible
Born-Oppenheimer molecular dynamics, Eq.\ (\ref{TRL}) \cite{ANiklasson06} and $iii)$ the proposed symplectic integration scheme,
Eqs. (\ref{SR}) and (\ref{SI}), with the optimal 4th order coefficients by McLachlan and Atela \cite{McLachlan92},
with $\kappa =4.617$.  In scheme $i)$ and $ii)$ the nuclear coordinates are integrated by the velocity-Verlet algorithm \cite{WCSwope82}.
Three SCF fixed point iterations without mixing were used in each force calculation.
This example clearly illustrates three levels of performance.
The linear interpolation scheme has a rapid decay, whereas
no systematic energy drift is seen for the time-reversible and the symplectic integration schemes. 
Most importantly, the higher-order symplectic approach improves the numerical accuracy, as measured by the amplitude 
of the energy fluctuations, by over an order of magnitude ($\sim$ 1/40) compared to the previous
time-reversible scheme, at the same level of computational cost.
Similar improvements are found in Tab.\ \ref{tab:DE}.

\begin{table}[t]
  \centering
  \caption{\protect Estimated total energy fluctuation amplitudes for the linear time-reversible (A) \cite{ANiklasson06}
    and the optimized 4th order symplectic (B) integration in Eqs.\ (\ref{SR}) and (\ref{SI}) \cite{McLachlan92}.  
    (3 SCF/force calculation)}\label{tab:DE}
  \begin{tabular}{cll}
   System   & A  ($\delta t = 0.5$ fs) &  B ($\delta t = 2.0$ fs) \\
    \hline
     (H$_2$O)$_{10}$ (RHF/3-21G) & 50 $\mu$Hartree    & 2 $\mu$Hartree \\
    \hline
     C$_2$F$_4$ (RHF/3-21G) & 80 $\mu$Hartree    & 2 $\mu$Hartree \\
    \hline
     F$_2$      (RHF/6-31G) & 20  $\mu$Hartree   & 0.07 $\mu$Hartree \\
  \end{tabular}
\end{table}

In summary, a Lagrangian formulation of time-reversible Born-Oppenheimer molecular dynamics was proposed, where
extended auxiliary electronic degrees of freedom evolve as a harmonic oscillator centered around the adiabatic propagation of the
self-consistent electronic ground state. The Lagrangian formulation enables the application of highly efficient
and accurate symplectic or geometric integration methods that are shown to be stable under incomplete SCF convergence.
For example, using a 4th-order symplectic integration scheme it was demonstrated how the accuracy is improved 
by over an order of magnitude compared to previous formulations.  With a Lagrangian formulation it is also possible 
to extend microcanonical simulations to other ensembles, for example, to Nos\'{e}-Hoover thermostats or Langevin dynamics.
The extended Born-Oppenheimer molecular dynamics presented here, which is based on a density matrix representation,
may thus open the door to alternative and even more versatile formulations of molecular dynamics. 

We gratefully acknowledge the support of the US Department of Energy through the LANL LDRD/ER program for this work.
Discussions with N. Bock, M. Challacombe, E. Chisholm, C. Greeff, W. Harrison, E. Holmstr\"{o}m, J. Hutter, S. Niklasson,
A. Odell, T. Peery, E. Rubensson, C.J. Tymczak, V. Weber, and J. Wills are gratefully acknowledged.

\end{document}